\documentclass[10pt]{article}
\usepackage{authblk}
\usepackage{graphicx}
\usepackage{amsfonts}
\usepackage{float}
\usepackage{mathtools}
\usepackage[backend=bibtex]{biblatex}
\usepackage{amsmath}
\usepackage{array}
\usepackage{subcaption}
\usepackage{algorithm}
\usepackage{algorithmic}

\usepackage{amssymb,latexsym,amsmath,amscd,slashed,mathtools,mathrsfs,bm,stmaryrd,dsfont,amsthm}
\usepackage{setspace,scalerel,color,graphicx,enumerate}

\newcommand{\ostar}{\mathbin{\mathpalette\make@circled\star}}

\usepackage{authblk}
\theoremstyle{definition}
\newtheorem{theorem}{Theorem}
\newtheorem{corollary}[theorem]{Corollary}
\newtheorem{lemma}[theorem]{Lemma}
\newtheorem{prop}[theorem]{Proposition}

\theoremstyle{definition}
\newtheorem{definition}{Definition}
\newtheorem{remark}[definition]{Remark}

\DeclareMathOperator{\Tr}{Tr}
\DeclareMathOperator{\stab}{Stab}
\DeclareMathOperator{\gr}{{\displaystyle \mathbf{Gr}}}
\DeclareMathOperator{\Gre}{GrE}

\DeclareMathOperator{\ex}{\mathbb{E}}
\DeclareMathOperator{\Wg}{\operatorname{Wg}}

\DeclareMathOperator{\Sym}{\mathfrak{S}}
\DeclareMathOperator{\unitary}{\mathrm{U}}

\newcommand{\ket}[1]{\left|#1\right\rangle}
\newcommand{\bra}[1]{\left\langle #1\right|}

\theoremstyle{plain}

\makeatletter

\date{} \makeatother
\title{The moments and asymptotic spectral distribution of the Grassmannian ensemble}
\newcommand{\authoraffiliation}[3]{%
  \par\addvspace{0.7\baselineskip}%
  \noindent\begin{minipage}{\linewidth}%
    \small\textbf{#1}\\
    #2\\
    \textit{E-mail address}: \texttt{#3}%
  \end{minipage}\par}
\AtEndDocument{%
  \section*{Author affiliations}
  \authoraffiliation{Aryan Gadhave}{Department of Mathematics, University of Cambridge, United Kingdom}{ag2381@cam.ac.uk}
  \authoraffiliation{Asghar Ghorbanpour}{Department of Mathematics, University of Western Ontario, Canada}{aghorba@uwo.ca}
  \authoraffiliation{Liam Hickey}{Department of Mathematics, University of Western Ontario, Canada}{lhickey8@uwo.ca}
  \authoraffiliation{Masoud Khalkhali}{Department of Mathematics, University of Western Ontario, Canada}{masoud@uwo.ca}
  \authoraffiliation{Allison Klingler}{Department of Mathematics, Amherst College, USA}{aklingler27@amherst.edu}
  \authoraffiliation{Nathan Pagliaroli}{Department of Combinatorics and Optimization, University of Waterloo, Canada}{npagliar@uwaterloo.ca}
  \authoraffiliation{Andrew Parks}{Department of Mathematics, Vanderbilt University, USA}{Andrew.parks@vanderbilt.edu}
}

\author{Aryan Gadhave, Asghar Ghorbanpour, Liam Hickey, Masoud Khalkhali, Allison Klingler, Nathan Pagliaroli, Andrew Parks}

\begin{document}
	\maketitle
	\begin{abstract}
		In this paper, we study random density matrices constructed by taking partial traces of projections onto random subspaces of a bipartite Hilbert space $\mathcal{H}_A \otimes \mathcal{H}_{B}$ distributed according to the pushforward measure of the action of the unitary group on the Grassmannian $\gr_{n}(\mathcal{H}_A \otimes \mathcal{H}_{B})$. This ensemble models a `typical' quantum subspace. All tracial moments of the density operator, as well as mixed tracial moments of reduced density operators, are computed explicitly using Weingarten calculus. There are several consequences of these results. In particular, we prove that the empirical spectral distribution of the density operator converges almost surely  to a rescaled Marchenko-Pastur law. We also prove that this ensemble can be written as a sum of asymptotically free random density operators, leading to an alternative proof of the convergence of the spectral distribution. Finally, we compute the asymptotic expansion of the average entanglement entropy.
	\end{abstract}

\section{Introduction}

Motivated by developments in quantum information theory, studying the typical
spectral and information-theoretic properties of finite-dimensional quantum
systems has become the focus of many directions of mathematical research. Several choices of probability measures on the mixed states of finite-dimensional Hilbert spaces have been studied using random matrix theory \cite{braunstein1996geometry,hall1998random,sommers2004statistical,bengtsson2017geometry}. In particular,  consider the pushforward measure of  the unique unitarily invariant probability measure on the pure states of the composite system $\mathcal{H}$  under the partial trace map to the state space of a subsystem
\begin{equation*}
    |\psi\rangle\mapsto \Tr_B(|\psi\rangle\langle \psi |).
\end{equation*}
 We will refer to the state space of $\mathcal{H}_A$ with this probability measure as the \textit{induced ensemble}. An exact formula for the von Neumann entropy of the induced ensemble was conjectured by Page \cite{page1993}, with the help of estimates obtained by Lubkin \cite{lubkin1978}. Page's conjecture was proved in \cite{foong1994proof,sanchez1995simple,sen1996average} using different techniques. 
These results have become a cornerstone in quantum information theory, black hole physics, and the study of thermalization in quantum systems \cite{harlow2016jerusalem,chen2022quantum,vardhan2022bound,chen2021observation}. 

Various probabilistic aspects of the induced ensemble have been studied. Its moments are related to those of the complex Wishart-Laguerre ensemble  \cite{nechita2007asymptotics} and the Ginibre ensemble \cite{zyczkowski2011generating}. In this work, we adapt these approaches to the study of `typical' quantum subspaces.  Let
$d_A=\dim_{\mathbb C}\mathcal H_A$, $d_B=\dim_{\mathbb C}\mathcal H_B$,
and $m=d_A d_B$. For a random $n$-dimensional subspace
$\mathcal K\subset\mathcal H_A\otimes\mathcal H_B$, we consider
\[
\mathcal{K} \mapsto\frac1n\Tr_B(P_{\mathcal K}),
\]
for $1\leq n\leq m$, where $P_{\mathcal K}$ is the orthogonal projection onto $\mathcal K$. The random subspace is distributed according to the pushforward measure of the Haar measure on the group of unitaries of $\mathcal H_A\otimes\mathcal H_B$. We call the distribution of the above random mapping the Grassmannian ensemble. Notions of `typical subspaces' are often used in proofs of coding theorems in quantum information theory  \cite{schumacher1995quantum}; as such, the Grassmannian ensemble should lend itself to such proofs. A typical quantum subspace can be thought of as analogous to a typical set in classical information theory. Similar questions posed for quantum subsystems are applicable to quantum subspaces, especially concerning entropy, data compression, and entanglement.

In this work, we will focus on studying the eigenvalue distribution and entropy of the Grassmannian ensemble. All work that computes the moments and entropies of the induced ensemble relies heavily, often implicitly, on its relationship with the Wishart-Laguerre ensemble and the application of orthogonal polynomial techniques. For the Grassmannian ensemble, a similar approach is not obvious,  so we use different tools: Weingarten calculus and asymptotic freeness. We note that the Grassmannian ensemble directly generalizes the induced ensemble, so these approaches can also be seen as alternative methods for studying the induced ensemble.

This article is organized as follows. 
In Section 2, we will compute the tracial moments of the Grassmannian ensemble using Weingarten calculus. 
These results then allow us to compute the limiting eigenvalue distribution of the Grassmannian ensemble in Section 3 via the method of moments. 
We further show using techniques of free probability theory that the reduced density operators are asymptotically free, which also provides an alternative proof of convergence to the limiting eigenvalue distribution. We also compute entanglement entropy using these results. In the Appendix, we include Monte Carlo simulations. 


\section{Grassmannian ensemble}

Consider the bipartite Hilbert space $\mathcal{H}=\mathcal{H}_A\otimes \mathcal{H}_B$ with complex dimensions $d_A$, $d_B$, and $m=d_Ad_B$, respectively. 
Throughout this paper, we will only consider finite-dimensional complex Hilbert spaces. The complex Grassmannian of $n$-dimensional subspaces of $\mathcal{H}$, which we will denote by $\gr_{n}(\mathcal{H})$ or simply as $\gr(n,m)$, admits a unique $\unitary(\mathcal{H})$-invariant probability measure. 
In fact, the unitary group $\unitary(\mathcal{H})$ acts transitively on $\gr_{n}(\mathcal{H})$. 
For any base point $\mathcal{K}_0\in \gr_n(\mathcal{H})$, the stabilizer subgroup $\mathrm{stab}(\mathcal{K}_0)=\unitary(\mathcal{K})\times \unitary(\mathcal{K}_0^\perp)$ is compact; therefore, the map 
\begin{equation}\label{themap}
    \phi:\unitary(\mathcal{H})\to \gr_n(\mathcal{H}), \qquad U\mapsto U(\mathcal{K}_0),
\end{equation}
induces a homeomorphism from $\unitary(\mathcal{H})/\mathrm{stab}(\mathcal{K}_0)$ to  $\gr_{n}(\mathcal{H})$. 
The pushforward of the Haar probability measure $\mu_{H}$ on $\unitary(\mathcal{H})$ under the map $\phi$ is a $\unitary(\mathcal{H})$-invariant probability measure on $\gr_n(\mathcal{H})$. 
Since the stabilizer group is compact, this probability measure, denoted by $\mu_{H}$, is the unique $\unitary(\mathcal{H})$-invariant probability measure on the homogeneous space $\gr_n(\mathcal{H})$ (see, e.g. \cite[Section 2.6]{Folland16}).
The independence of our construction from the choice of the base subspace $\mathcal{K}_0$ follows from this uniqueness, which is closely related to $\unitary(\mathcal{H})$-invariance of the Haar measure.

Next, we assign a quantum state on the subsystem $\mathcal{H}_A$ to any element of $\gr_n(\mathcal{H})$.  
For any $\mathcal{K} \in \gr_n(\mathcal{H})$, we denote the orthogonal projection onto $\mathcal{K}$ by $P_{\mathcal{K}} \in \mathcal{H}$.
We then use the partial trace to map this projection to the state space $\mathcal{S}(\mathcal{H}_A)$ of $\mathcal{H}_A$. The partial trace is defined on $B(\mathcal{H})$ and models a quantum-physical setting in which a state in the bipartite system $\mathcal{H}_A \otimes \mathcal{H}_B$ is examined only using observables in the subsystem $A$.
It is a linear map uniquely characterized by the following property:
\begin{equation*}
    \Tr\left((M\otimes \mathbb{I}) T\right)=\Tr(M\Tr_B(T)),\qquad T\in B(\mathcal{H}),\, M\in B(\mathcal{H}_B).
\end{equation*}
 By tracing out the subsystem $B$ from the projection $P_{\mathcal{K}}$, we obtain a non-negative operator on $\mathcal{H}_A$ whose trace is equal to the dimension of $\mathcal{K}$ and must therefore be normalized:
$$\Tr(\Tr_B(P_{\mathcal{K}}))=\Tr(P_{\mathcal{K}})=\dim(\mathcal{K})=n.$$ 
In summary, we have the map $\pi:\gr_n(\mathcal{H})\to \mathcal{S}(\mathcal{H}_A)$,
\begin{equation}\label{equ: GRQS}
\mathcal{K}\mapsto \frac{1}{n}\Tr_B(P_{\mathcal{K}}).
\end{equation}
We use the maps $\pi$ and $\phi$ to push forward the $\unitary(\mathcal{H})$-invariant measure $\mu_H$ to the state space $\mathcal{S}(\mathcal{H}_A)$.

\begin{definition}
The \textit{Grassmannian ensemble} of a subsystem $\mathcal{H}_A$ of the bipartite system $\mathcal{H}=\mathcal{H}_A\otimes \mathcal{H}_B$ with parameters $(n,d_A;d_B)$ is the set of all quantum states $\mathcal{S}(\mathcal{H}_A)$ of $\mathcal{H}_A$  with the pushforward measure $\pi_* \phi_{*}\mu_{H}$.  
 We will denote this distribution by $\rho \sim \Gre (n,d_A;d_B)$.
\end{definition}

\begin{remark}
    The measure $\pi_*\phi_{*}\mu_{H}$ is a $\unitary(\mathcal{H}_A)$-invariant measure. 
    Let $U\in \unitary(\mathcal{H}_A)$. Then the unitary operator $U\otimes \mathbb{I}_B$ is in $\unitary(\mathcal{H})$ and we have 
$$U \Tr_B(P_{\mathcal{K}}) U^*= \Tr_B\left((U\otimes \mathbb{I}_B) P_{\mathcal{K}}(U\otimes \mathbb{I}_B)^*\right)=\Tr_B(P_{(U\otimes \mathbb{I}_B) \mathcal{K}}),\qquad  U\in \unitary(\mathcal{H}_A).$$
In other words,
$U\pi(\mathcal{K}) U^*=\pi((U\otimes \mathbb{I}_B)\mathcal{K})$, and the $\unitary(\mathcal{H})$-invariance of $\mu_H$ completes the proof of $\unitary(\mathcal{H}_A)$-invariance of $\pi_* \phi_{*}\mu_{H}$. 
\end{remark}

To understand this ensemble, we wish to find its eigenvalue distribution. This is an example of the quantum marginal problem. We approach this problem via the method of moments. The integration in the calculation of the tracial moments of the Grassmannian ensemble can be turned into an integral on the unitary group $\unitary(\mathcal{H})$. 
In fact, fix any base point $\mathcal{K}_0$, and let $U$ be a Haar-distributed random unitary matrix. Then we have   
$$\frac{1}{n}\Tr_B(UP_{\mathcal{K}_0}U^*)\sim \Gre (n,d_A;d_B).$$
Therefore, we may write 
\begin{align*}
\ex\left[\Tr\left(\rho^k\right) \right]&
=\int_{\gr_n(\mathcal{H})} \Tr\left(\left(\frac{1}{n}\Tr_B(P_{\mathcal{K}})\right)^k\right) ~ d\phi_{*}\mu_{H}(\mathcal{K})\\
&=\int_{\unitary(\mathcal{H})} \Tr\left(\left(\Tr_B\left(U{\small\frac{1}{n}}P_{\mathcal{K}_0}U^*\right)\right)^k\right) ~ d\mu_{H} (U).
\end{align*}
If $\mathcal{K}$ and a basis for $\mathcal{H}$ are carefully chosen, then the above integral can be expressed in terms of integrals of products of unitary-matrix entries of the following form:

\begin{equation}\label{momentsofUn}
    \int_{\unitary(m)}U_{i_1j_1}\dots U_{i_kj_k}\overline{U_{i_1'j_1'}}\dots \overline{U_{i_{k'}'j_{k'}'}} ~ d\mu(U).
\end{equation}
These integrals are called moments of the unitary group or Weingarten integrals. In the next section, we will review the theory required to evaluate these integrals.

Alternatively, the  unitary invariance of the induced measure can be used to turn this ensemble into an ensemble of density matrices in the following way. 
The state $\frac{1}{n}\Tr_B(P_{\mathcal{K}})$ is of the form $\Tr(\cdot \, \rho_{\mathcal{K}})$ for some density operator $\rho_{\mathcal{K}}$. 
By fixing an orthonormal basis for $\mathcal{H}_A$, it can be replaced by its matrix representation in that basis. 
Since all the operations defined above are $\unitary(\mathcal{H})$-invariant and hence $\unitary(\mathcal{H}_A)$-invariant, we can safely see the map $\phi$ from $\gr_n({\mathcal{H}})$ to the space of density matrices on $\mathcal{H}_A$. Given the projection $P_{\mathcal{K}_{0}}$ onto the subspace $\mathcal{K}_0$, we can decompose its conjugation into a sum of one-dimensional projections $\ket{\phi_i}\bra{\phi_{i}}$ for an orthonormal basis $\{\ket{\phi_i}\}_{i}$ for $\mathcal{H}$. In conclusion, the random variable we are interested in can be written as
\begin{equation*}
    \frac{1}{n}\Tr_{B}(U P_{\mathcal{K}_{0}}U^{*}) = \frac{1}{n}\sum_{i=1}^{n} \Tr_{B}(U \ket{\phi_i}\bra{\phi_{i}}U^{*}). 
\end{equation*}

\begin{definition}
Let $\{\ket{\phi_i}\}$ be an orthonormal set of vectors in $\mathcal{H}_A\otimes \mathcal{H}_B$.  
Define the family $\{\rho_i\}$ by 
\begin{equation*}
\rho_i=\Tr_B(U\ket{\phi_i}\bra{\phi_i}U^*),
\end{equation*}
where  $U$ is a Haar-distributed random unitary matrix. We refer to $\{\rho_{i}\}_{i}$ as a set of \textit{orthogonally induced random quantum states}. 
\end{definition}

\begin{remark}
    Note that a Grassmannian ensemble can be defined for any set of orthogonal projections, not necessarily one-dimensional ones. We will restrict our attention in this work to only one-dimensional projections.
\end{remark}

  The Grassmannian ensemble with parameter $n=1$ reduces to the induced ensemble defined in the Introduction. Given any two unit vectors $\phi$ and $\phi'$, there exists a unitary operator $U_0$ such that $U_0\phi=\phi'$.
This implies that
$$\Tr_B(U\ket{\phi'}\bra{\phi'}U^*)=\Tr_B(UU_0\ket{\phi}\bra{\phi}(UU_0)^*).$$
However, if $U$ is Haar-distributed, then both $U$ and $UU_0$ are Haar-random unitary matrices. 
Therefore, the states in an orthogonally induced family are identically distributed with the same distribution as the induced ensemble, which is known explicitly. An integral representation of the joint distribution may be obtainable following the approach in \cite{collins2023projection}.

\subsection{Weingarten calculus and tracial moments}

We will briefly review the Weingarten calculus and introduce the notation that we will use in this section.
For a detailed account of the topic, we refer the reader to \cite{collins2022weingarten}.

The integrals of the form \eqref{momentsofUn} were first studied in the theoretical physics literature; see, e.g., \cite{UllahPorter63}.
Benoît Collins \cite{collins2003moments} was the first to provide an explicit formula to compute them.
This approach was later generalized for other classical groups in \cite{CollinsSniady06}.
The collection of results and techniques related to the evaluation of these integrals is now known as the {\it Weingarten calculus}.
The term was selected by Collins in honor of Donald Weingarten, who studied the asymptotics of unitary polynomial integrals \cite{weingarten1978asymptotic}.

A partition $\pi$ (with $l$ blocks) of a positive integer $k$, denoted by $\pi \vdash k$, is a tuple $\pi= (\pi_1,\dots,\pi_l)$ of positive integers such that $\sum_{j=1}^l \pi_j=k$. 
For any partition $\pi\vdash k$, we define the polynomial $$C_\pi(z)=\prod_{i=1}^{l}\prod_{j=1}^{\pi_i}(z+j-i).$$
The cycle structure of a permutation $\sigma \in \Sym_k$ defines a partition of $k$ consisting of the lengths of the cycles of $\sigma$. The number of blocks in this partition is simply the number of cycles in $\sigma$, which we will denote by $\#(\sigma)$.
For any partition $\pi\vdash k$, the (irreducible) character $\chi^\pi$ is the character of the irreducible representation indexed by $\pi$.

The \textit{(unitary) Weingarten function} $\Wg$ is a complex-valued function defined as
\begin{equation*}
    \Wg(\sigma,z) = \frac{1}{k!}\sum_{\substack{\pi \vdash k\\ C_\pi(z)\not= 0}}\frac{\chi^{\pi}(\text{id}_{k})\chi^{\pi}(\sigma)}{C_\pi(z)},\qquad z\in \mathbb{C},\sigma\in \Sym_k.
\end{equation*}
The fundamental theorem of Weingarten calculus expresses the value of Weingarten integrals \eqref{momentsofUn} in terms of the Weingarten function. 

To keep our notation concise, we will take advantage of the multi-index notation and use the following conventions: 
$$\begin{matrix}
\begin{matrix}
 U_{\mathbf{i}\mathbf{j}}=U_{i_1j_1}\dots U_{i_kj_k},\\
\delta_\sigma(\mathbf{i},\mathbf{j})=\prod\limits_{t=1}^{k}\delta_{i_{\sigma^{-1}(t)}j_t},
\end{matrix}
 & \qquad \mathbf{i}=(i_1,\dots,i_k),\mathbf{j}=(j_1,\dots,j_k)\in \{1,\dots,m\}^k,\sigma\in \Sym_k.
\end{matrix}$$

\begin{theorem}\cite{CollinsSniady06}
 Let $U$ be a Haar-distributed random unitary matrix of size $m$. Then
    \begin{equation*}\label{FTWC}
\ex \Big[ U_{\mathbf{i}\mathbf{j}}\overline{U_{\mathbf{i}'\mathbf{j}'}} \Big]=
\begin{cases}
\displaystyle \sum_{\sigma,\tau\in \Sym_k } \delta_\sigma(\mathbf{i},\mathbf{i}')\delta_\tau(\mathbf{j},\mathbf{j}')\Wg(\sigma^{-1}\tau,m) & \text{ if } k=k',\\ &\\
0 & \text{ if } k\neq k',
\end{cases}
\end{equation*}
where  $\mathbf{i,j}\in\{1,\dots,m\}^k$ and $\mathbf{i',j'}\in\{1,\dots,m\}^{k'}$.
\end{theorem}

Note that the theorem essentially provides a formula for the integral of the entries of $U^{\otimes k}\otimes (U^*)^{\otimes k}$ and can equivalently be expressed in the form of an operator-valued integral.
To see this, we need to first fix the action of the symmetric group $\Sym_k$ on $\mathcal{H}^{\otimes k}$ defined by permuting the tensor factors. 
More precisely, for any permutation $\sigma \in \Sym_k$, and for vectors $v_j\in\mathcal{H}$, the action is defined by
\begin{equation}\label{equ: actionofsymontensorHilbert}
    \tau(v_1\otimes v_2 \otimes\dots \otimes v_k) = v_{\tau^{-1}(1)}\otimes v_{\tau^{-1}(2)}\otimes\dots\otimes v_{\tau^{-1}(k)}.
    \end{equation}
Note that $\delta_\sigma(\mathbf{i},\mathbf{j})$ are the entries of the matrix representation of the action of $\sigma\in\Sym_k$ on $\mathcal{H}^{\otimes k}$ in any orthonormal basis of the form $\{e_{i_1}\otimes \cdots\otimes  e_{i_k}|\mathbf{i}\in\{1,\dots,m\}^k\}$.
Therefore, we have (see, e.g., \cite{CollinsSniady06}):  
\begin{equation}\label{equ: AlternativeFTWC}
\ex\left[U^{\otimes k} O (U^*)^{\otimes k}\right]=\sum_{\sigma,\tau\in \Sym_k } \Wg(\sigma^{-1}\tau,m)\Tr(\tau^{-1} O)\, \sigma.
\end{equation}

The trace term on the right hand side of equation \eqref{equ: AlternativeFTWC} can be further simplified when the given operator \(O\) is the tensor product of operators $A_j$ from  $B(\mathcal{H})$. 
 Let $\gamma$ be the full cycle $(1\,2\,\dots\,k)\in\Sym_k$. Then we have 
\begin{equation}\label{equ: swaptrick-original}
\Tr(\gamma (A_1 \otimes A_2\otimes \dots \otimes A_k)) = \Tr(A_k\cdots A_2A_1).
\end{equation} 
Note that indices belonging to distinct cycles are summed independently, and for a given cycle $(i_{1} \cdots i_{c})$ the associated term is $\Tr(A_{i_{i}}\cdots A_{i_{k}})$. This allows one to deduce the following result, commonly referred to as the ``swap trick''.

\begin{lemma}\label{lem: simplelemma}
    Let $A_t$, for all $t=\{1,\dots,k\}$, be operators on the Hilbert space $\mathcal{H}$. Then 
\begin{equation}
    \Tr\Big(\tau (A_1 \otimes A_2\otimes \dots \otimes A_k) \Big)=\prod_{c \text{ is a cycle of }\tau} \Tr(A_c),\quad \tau\in\Sym_k.
\end{equation}
Here, for any cycle $c=(i_1\, i_2\, \dots \, i_l)$, we use the notation $A_c=A_{i_l}A_{i_{l-1}}\dots A_{i_1}$.
\end{lemma}

The case where all $A_j$'s are the same operator $A\in B(\mathcal{H})$, in other words,  $O=A^{\otimes k}$,
 is studied in \cite[Theorem 3.1]{collins2014integration}.
It is clear that for a permutation $\tau \in \Sym_k$ with cycle structure $(c_1, c_2, \dots, c_l)$, the term $\Tr_{\tau}(A)$ is defined as follows:
\begin{equation*}
\Tr_{\tau}(A) \coloneqq \prod_{j=1}^l \Tr\!\left(A^{c_j}\right).
\end{equation*}
Furthermore, the effect of the presence of the partial trace was carefully analyzed in \cite{matsumoto2023moments} where the random quantum marginal problem is explored using Weingarten calculus.
What follows is a special case of \cite[Theorem 2.3]{matsumoto2023moments}.

\begin{prop}\label{prop: frommatsumoto2023moments}
Let $U$ be a Haar-distributed random matrix in $\unitary(m)$, and let $\mathcal{H}_A = \mathbb{C}^{d_A}$ and $\mathcal{H}_B = \mathbb{C}^{d_B}$, with $m = d_A d_B$. 
For any real diagonal matrix $\Lambda = \operatorname{diag}(\lambda_1, \dots, \lambda_m)$, we have
\[
\ex\!\left[\prod_{t=1}^{k} \Tr_{B}(U\Lambda U^*)_{i_t,j_t}\right]
= \sum_{\sigma,\tau\in \Sym_{k}} \delta_{\sigma}(\mathbf{i},\mathbf{j})\, d_B^{\#(\sigma)}\,
\Wg\!\left(\sigma^{-1}\tau,m\right)\, \operatorname{Tr}_{\tau}(\Lambda).
\]

\end{prop}
Note that the above formula is valid not only for any diagonal matrix $\Lambda$, but also for a general Hermitian matrix.
This is because both sides of the equality are invariant under conjugation by a unitary matrix $V \in \unitary(m)$: the left side, because of the $\unitary(m)$-invariance of the Haar measure on $\unitary(m)$, and the right side, because of the tracial property of $\Tr$ and the fact that conjugation by $V$ is a group homomorphism.
Therefore, the equality holds for the normalized projection $\Lambda = \frac{1}{n}P_{\mathcal{K}}$.\\
Moreover, the projection $P_{\mathcal{K}}$ is an idempotent with trace $\dim \mathcal{K} = n$.
Then, for any $\tau \in \Sym_k$ with cycle structure $(c_1, c_2, \dots, c_l)$, we have
\begin{equation}\label{equ: tracetauofprojection}
\Tr_{\tau}\left(\frac{1}{n}P_{\mathcal{K}}\right)
=\prod_{j=1}^l \Tr\left(\big(\frac{1}{n}P_{\mathcal{K}}\Big)^{c_j}\right)  =\prod_{j=1}^l \frac{1}{n^{c_j}}\Tr(P_{\mathcal{K}}^{c_j})=\prod_{j=1}^l n^{1-c_j}=n^{\#(\tau)-k}.
\end{equation}

Now, we can  compute the tracial moments of the Grassmannian ensemble.

\begin{theorem}\label{thm: tracial moments}
For any state $\rho\sim \Gre (n,d_A;d_B)$, the tracial moments are given by 
\begin{equation*}
\ex\left[\Tr(\rho^k)\right]=\sum\limits_{\sigma, \tau \in \Sym_k}\Wg(\sigma^{-1}\tau,d_Ad_B)d_A^{\#(\gamma^{-1}\sigma)}d_B^{\#(\sigma)}n^{\#(\tau)-k}.
\end{equation*}
Here $\gamma$ is the $k$-cycle $(1\,2\, \dots\, k)\in\Sym_k$. 
\end{theorem}
\begin{proof}
It is enough to prove the result for the Grassmannian random quantum state $\rho$ defined by \eqref{equ: GRQS}.\\
The multi-indices $\mathbf{i}, \mathbf{j} \in \{1, \dots, d_A\}^k$ that appear in the formula given by Proposition \ref{prop: frommatsumoto2023moments}, when writing the trace of $\rho^k$ in terms of the entries of $\rho$, are related by $i_t = j_{\gamma(t)}$, where $\gamma$ is the full cycle $(1\,2\,\dots\,k) \in \Sym_k$. 
Denote this multi-index $\mathbf{j}$ by $\gamma(\mathbf{i})$.

Then Proposition \ref{prop: frommatsumoto2023moments}, together with \eqref{equ: tracetauofprojection}, gives the following:
\begin{align*}
\ex\left[\Tr(\rho^k)\right] &=\sum\limits_{\mathbf{i}}\ex\left[\rho_{i_1,i_2}\,\rho_{i_2,i_3}\,\dots \rho_{i_t,i_1}\right]\\
&=\sum\limits_{\mathbf{i}}\sum\limits_{\sigma, \tau \in \Sym_k}\delta_{\sigma}(\mathbf{i},\gamma(\mathbf{i}))\,d_B^{\#(\sigma)}\Wg(\sigma^{-1}\tau,d_Ad_B)\,n^{\#(\tau)-k} \\
    &= \sum_{\sigma, \tau \in \Sym_k} \left( \sum_{\mathbf{i}} \delta_{\sigma}(\mathbf{i}, \gamma(\mathbf{i})) \right) d_B^{\#(\sigma)} \Wg(\sigma^{-1}\tau, d_Ad_B)\, n^{\#(\tau)-k}.
\end{align*}
To evaluate the inner sum, we are effectively counting the number of multi-indices $\mathbf{i}$ such that
$$i_{\sigma(t)}=j_t\Leftrightarrow i_{\sigma(t)}=i_{\gamma(t)}\Leftrightarrow 
i_{\gamma^{-1}\sigma(t)}=i_{t}, \qquad t\in \{1,2,\dots,k\}. 
$$
This invariance under permutations of the entries can be extended further to show that the multi-index must remain invariant under any power of the permutation $\gamma^{-1}\sigma$.
As a result, the entries of the multi-index are constant on all positions that can be mapped to each other by any power of the permutation $\gamma^{-1}\sigma$. In other words, the entries of the multi-index are constant on the orbits of the subgroup $\langle\gamma^{-1}\sigma\rangle$ acting on $\{1,\dots,k\}$.
Therefore, the multi-indices that we are trying to count are those that are constant on the positions corresponding to the cycles of $\gamma^{-1}\sigma$.
Since each constant has $d_A$ possible values for each cycle, the sum evaluates to:
\begin{equation*}
   \sum_{\mathbf{i}} \delta_{\sigma}(\mathbf{i}, \gamma(\mathbf{i})) =\sum_{\mathbf{i}} \delta_{\gamma^{-1}\sigma}(\mathbf{i}, \mathbf{i})  = d_A^{\#(\gamma^{-1}\sigma)}.
\end{equation*}
Applying this identity  gives our result.
\end{proof}

As an example, the second tracial moment can be calculated for general $m=d_Ad_B$ and $n$:
\begin{equation}\label{equ: secondtracialmoment}
\begin{aligned}
     \ex [\Tr \rho^2] &=
     \Wg\big((1)(2),m\big)\, d_A \, d_B^2+\Wg\big((1\,2),m\big)\,d_A^{2}\, d_B\\   &\,\,\,+\Wg\big((1\,2),m\big)\,d_A\,d_B^{2}\,n^{-1}+\Wg\big((1)(2),m\big)\, d_A^{2}\, d_B \, n^{-1}\\
     &=\frac{d_A d_B^2}{m^2-1}-\frac{d_A^{2}d_B}{m(m^2-1)} -\frac{d_Ad_B^{2}}{m(m^2-1)n}+\frac{d_A^{2}d_B}{(m^2-1)n}\\
     &=\frac{d_A(m-n)+d_B(m n-1)}{\left(m^2-1\right)n}\\
     &=\frac{d_A^2 d_B+d_Ad_B^2 n-d_A n-d_B}{(d_A^2d_B^2-1)n}.
\end{aligned}
\end{equation}
One can find the third tracial moment through similar calculations.
\begin{equation*}
    \ex [\Tr \rho^3]= 
    \frac{(m-n)(3m^2n-5m-2n)+d_A^2(m-2n)(m-n)+d_B^2(m^2n^2-3mn-2n^2+4)}{(m^2-4)(m^2-1)n^2}.
\end{equation*}
Setting $n = 1$, we recover the tracial moments of the induced ensemble, which can be found in the literature; see, e.g., \cite{nechita2007asymptotics}.
\begin{equation}\label{equ: tracialmomentsofinduced}
    \ex [\Tr \rho^2]=\frac{d_A+d_B}{m+1},\qquad\text{ and }\qquad \ex [\Tr \rho^3]=\frac{d_A^2+d_B^2+3m+1}{(m+1)(m+2)}.
\end{equation}

\subsection{Mixed tracial moments}\label{sec: mixed moments}

In this section, we shall compute the mixed tracial moments of the orthogonally induced random states. 
In the following section, we will use this result to prove that orthogonally induced random states are asymptotically free.

Let $\ell_{1},\ell_{2},...,\ell_{k} \geq 1$, and let $w$ be a word over the alphabet $\{\rho_{1},...,\rho_{n}\}$, considered up to cyclic rotations. Let $\iota(w)$ denote the cyclic sequence of state labels in the word; for example, $\iota(\rho_{1}^2 \rho_{2} \rho_{1}^{2} \rho_{3}^{4}) =(1,1,2,1,1,3,3,3,3)$. We then consider the following stabilizer group:
$$\stab(w) := \left\{\sigma \in \Sym_{\ell_{1}+...+\ell_{k}}| \forall 1\leq i\leq k, \iota(\sigma w) =\iota(w) \right\}.$$ 


\begin{theorem}\label{thm: Mixed tracial moments}
Let  $\frac{1}{n}\sum_{t=1}^{n} \rho_{t} \sim \Gre (n,d_{A};d_{B})$. Let $w$ be a word over the alphabet of states $\{\rho_{1},...,\rho_{n}\}$. Then 
\begin{equation*}
\ex\left[\Tr(w)\right]=\sum_{\substack{\sigma \in \Sym_{\ell} \\ \tau \in \stab(w)}} \Wg(\sigma, m) d_A^{\#(\gamma_{\ell}\sigma\tau)} d_B^{\#(\sigma\tau)},
\end{equation*}
where $\ell := \ell_{1}+...+\ell_{k}$.
\end{theorem}

\begin{proof}
Let $\{P_t\}$ denote the orthogonal projections onto the elements of an orthonormal basis of $\mathcal{H}$ and $U$ be a Haar-distributed unitary operator  such that 
$$\rho_t=\Tr_B(UP_t U^*).$$
We will denote $UP_tU^{*}$ by $Q_t$.
Consider the word $w=\rho_{t_1}^{\ell_1}\rho_{t_2}^{\ell_2}\dots\rho_{t_k}^{\ell_k}$, where no two neighbouring states are the same. We first use the swap trick to express our tracial moment as
\begin{align*}
\ex\left[\Tr(\rho_{t_1}^{\ell_1}\rho_{t_2}^{\ell_2}\dots\rho_{t_k}^{\ell_k})\right]
&= \ex [\Tr \gamma (\rho_{t_1}^{\ell_1}\rho_{t_2}^{\ell_2}\dots\rho_{t_k}^{\ell_k})]\\
&=\ex\left[\Tr\left(\Big(\bigotimes_{j=1}^{k} \Tr_B (Q_{t_j}^{\otimes \ell_j})\Big)\gamma^{A}_\ell\right)\right]\\
&=\ex\left[\Tr\left(\Tr_{\mathbf{B}} \Big(\bigotimes_{j=1}^{k} Q_{t_j}^{\otimes \ell_j}\Big)\gamma^{A}_\ell \right)\right]\\
&=\ex\left[\Tr\left( \Big(\bigotimes_{j=1}^{k} Q_{t_j}^{\otimes \ell_j}\Big)(\gamma^{A}_\ell\otimes\mathbb{I}_{\mathbf{B}})\right)\right]\\
&=\Tr\left( \ex\left[\bigotimes_{j=1}^{k} Q_{t_j}^{\otimes \ell_j}\right](\gamma^{A}_\ell\otimes\mathbb{I}_{\mathbf{B}})\right).
\end{align*}
Here $\gamma^{A}_\ell$ is the action of the cycle $(1\,2\,\dots \, \ell)\in \Sym_\ell$, for $\ell=\ell_1+\dots+\ell_k$, acting on $\mathcal{H}_A^{\otimes k}$ and $\mathbb{I}_{\mathbf{B}}$ is the identity of $\mathcal{H}_B^{\otimes l}$. 
Also, $\Tr_{\mathbf{B}}$ is the partial trace over $\mathcal{H}_B^{\otimes \ell}$ as a subsystem of $\mathcal{H}^{\otimes \ell}=\mathcal{H}_A^{\otimes \ell}\otimes \mathcal{H}_A^{\otimes \ell}$.
Using the alternative form of the fundamental theorem of Weingarten calculus \eqref{equ: AlternativeFTWC}, we can now evaluate the expected value: 
\begin{align*} 
\ex\left[\bigotimes_{j=1}^{k} Q_{t_j}^{\otimes \ell_j}\right] 
&= \ex\left[U^{\otimes \ell}\, \Big(\bigotimes_{j=1}^{k} P_{t_j}^{\otimes \ell_j}\Big)\,(U^{*})^{\otimes \ell}\right]= \sum\limits_{\sigma, \tau \in \Sym_{\ell}}\Wg(\sigma \tau^{-1},m)\Tr\left(\Big(\bigotimes_{j=1}^{k} P_{t_j}^{\otimes \ell_j}\Big)\tau\right)\sigma.
\end{align*}
To further analyze the trace term, we apply Lemma \ref{lem: simplelemma}. 
If a cycle $c$ includes two indices  that correspond to two different projections, then the product operator $P_{c}$ will be zero by orthogonality.
We can conclude that only permutations that do not mix up the different projections outside of a given cycle will contribute to this sum.
These permutations are precisely  $\stab(w)$.

Assume that $\tau\in\stab(w)$. Then, for any cycle $c$, the operator $A_c$ will be the product of several copies of only one projection. 
So, $\Tr(A_c)=\Tr(P_j^r)=\Tr(P_j)=1.$ 
Therefore, for any such permutation, we have 
$$\Tr\left(\Big(\bigotimes_{j=1}^{k} P_{t_j}^{\otimes \ell_j}\Big)\tau\right)=1.$$
Finally, we evaluate the following term:
$$\Tr(\sigma(\gamma_\ell^A \otimes \mathbb{I}_{\mathbf{B}}))
=\Tr(\sigma^A\gamma_\ell^A\otimes \sigma^B)
=\Tr(\sigma^A\gamma_\ell^A)\Tr(\sigma^B)
=\left(\sum_{\mathbf{i}}\delta_{\sigma\gamma_\ell}(\mathbf{i},\mathbf{i})\right) \left(\sum_{\mathbf{i}}\delta_{\sigma} (\mathbf{j},\mathbf{j})\right)=d_A^{\#(\gamma_\ell \sigma)}d_B^{\#(\sigma)}.$$
The last equality is exactly the identity established in the last paragraph of the proof of Theorem \ref{thm: tracial moments}.

Thus, we have
\begin{align*}
\ex\left[\bigotimes_{j=1}^{k} P_{t_j}^{\otimes \ell_j}\right] 
&= \sum_{\substack{\sigma \in \Sym_{\ell} \\ \tau \in \stab(w)}} \Wg(\sigma \tau^{-1},m)\Tr(\sigma(\gamma_\ell^A \otimes \mathbb{I}_{\mathbf{B}}))\\
&= \sum_{\substack{\sigma \in \Sym_{\ell} \\ \tau \in \stab(w)}}  \Wg(\sigma\tau^{-1},m)d_A^{\#(\gamma_\ell \sigma)}d_B^{\#(\sigma)} \\
&= \sum_{\substack{\sigma \in \Sym_{\ell} \\ \tau \in \stab(w)}} \Wg(\sigma, m) d_A^{\#(\gamma_\ell\sigma\tau)} d_B^{\#(\sigma\tau)}.
\end{align*}

\end{proof}

\begin{remark}
    We note that the above proof also works when the sizes of all the projections in the construction of the density matrix are not strictly one, in which case the resulting groups' sizes in the second summation would depend on the sizes of these projections in a straightforward manner.
\end{remark}
As an example, we can compute that 
\begin{align*}
\ex [\Tr(\rho_1 \rho_2)]&=\Wg((1)(2),m)d_A^{\#(12)}d_B^{\#((1)(2))}+\Wg((12),m)d_A^{\#((1)(2))}d_B^{\#((12))}\\
&=\frac{d_A d_B^2}{m^2-1}-\frac{d_A^{2}d_B}{m(m^2-1)} =\frac{md_B-d_A}{m^2-1}.
\end{align*}

\section{The limiting distribution of eigenvalues}
In this section, we will use the asymptotics of the Weingarten functions and the tracial-moment formula from the previous section to compute the limiting eigenvalue distribution of the Grassmannian ensemble. We will provide two distinct paths to proving that the eigenvalue distribution of properly normalized Grassmannian ensembles converges to the Marchenko-Pastur distribution. In the first method, we directly compute the moments of the Grassmannian ensemble for large system size using Theorem \ref{thm: tracial moments}, the asymptotics of the Weingarten function and properties of non-crossing permutations. Alternatively, we use tools from free probability theory and study the limit of mixed moments for large system size from Theorem \ref{thm: Mixed tracial moments}, showing that they are asymptotically free. Convergence to the Marchenko-Pastur law then follows from properties of the free cumulants.

\subsection{Non-crossing permutations}

 For a permutation $\pi \in S_{n}$ we denote by $\#(\pi)$ the number of cycles in $\pi$ and by $|\pi|$ the length of the permutation, i.e., the minimal number of transpositions $\pi$ can be written as. The number of cycles and length are related by $\# \pi + |\pi| =n$. Note that $\#(\pi^{-1}) = \#(\pi)$, so it follows that $|\pi| = |\pi^{-1}|$.
For any two permutations $\pi,\sigma\in S_n$, we have
$$\#(\pi)+\#(\pi^{-1}\sigma)+\#(\sigma) \leq  n +2.$$
The equality case is of particular interest to us. 

\begin{definition}
Let $\pi\in S_{n}$.
We say that the permutation $\pi$ is \textit{non-crossing}  if 
$$\#(\pi^{-1}\gamma_{n}) +\#(\pi)=n+ 1.$$
We denote the set of non-crossing permutations in $S_{n}$ as $\text{NC}(n)$.
\end{definition}

\begin{definition}\cite{mingo2004annular}
    For $p,q \ge1 $, let $\pi \in S_{p+q}$.  Define the permutation $\gamma_{p,q} :=(1\, 2\,\dots \,p)(p+1\, \dots \,p+q)$. Then we say $\pi$ is \textit{annular non-crossing} when 
    \begin{equation*}
        \#(\gamma_{p,q}\,\pi^{-1})+\#(\pi)= p+q+2.
    \end{equation*}
     We denote this set of non-crossing permutations by ${\text{NC}}(p,q)$.
\end{definition}

\subsection{Asymptotics}
Given a sequence of random $N$-by-$N$ matrices $X_{N}$ with eigenvalues $\{\lambda_{i}\}_{i=1}^{N}(X_{N})$, the \textit{empirical spectral measure } is defined as 
\begin{equation*}
    \mu_{X_{N}} :=\frac{1}{N}\sum_{i=1}^{N} \delta_{\lambda_{i}(X_{N})}.
\end{equation*}
In general, as the matrix size $N$ grows, this measure is used to describe the asymptotic eigenvalue distribution. Note that the empirical spectral measure is a random variable taking values in the space of probability measures that usually converges to a (deterministic) probability measure. More precisely, given a test function $\varphi \in C_{b}^{1}(\mathbb{R})$, we say that the empirical spectral measure $\mu_{X_{N}}$
converges in distribution to a probability measure $\mu$ if 
\begin{equation*}
    \lim_{N \rightarrow \infty}\ex \left[ \int \varphi d\mu_{X_{N}} \right] = \int \varphi d\mu.
\end{equation*}
We say it converges in probability if for any test function $\varphi \in C_{b}^{1}(\mathbb{R})$ and $\epsilon>0$ we have that 
\begin{equation*}
    \lim_{N \rightarrow \infty}\mathbb{P} \left[ \left|\int \varphi d\mu_{X_{N}} -\int \varphi d\mu\right|> \epsilon\right] =0.
\end{equation*}

For example, the empirical spectral measure of Wigner-type ensembles converges to the Wigner semicircular measure, and Wishart-Laguerre ensembles are known to converge to the Marchenko-Pastur probability measure. For some $c \in (0,\infty)$, the Marchenko-Pastur probability measure  is defined as  
$$d\mu^{MP}_{s}(x) = \max\{1-\frac1 s,0\}\delta_0+\frac{\sqrt{(b-x)(x-a)}}{2 \pi s x} \mathbf{1}_{ [a,b]}dx,$$
where $a = (1 - \sqrt{s})^2, b = (1 + \sqrt{s})^2.$ 
Its moments are 
\begin{equation*}
    \int x^{k} d\mu_{s}^{MP} =\sum_{\pi \in \text{NC}(k)}s^{k-\#(\pi)},
\end{equation*}
 see \cite{nica2006lectures}.
The Marchenko-Pastur probability measure appears in the limiting eigenvalue distribution of the induced ensemble \cite{nechita2007asymptotics}. It is therefore somewhat unsurprising that it appears in the context of the Grassmannian ensemble. Using the formula for the tracial moments from Theorem \ref{thm: tracial moments}, we will now compute the rescaled empirical spectral measure of eigenvalues for the  Grassmannian ensemble. We additionally need the leading-order asymptotics of the Weingarten function as $m$ tends to infinity, which from Weingarten's work \cite{weingarten1978asymptotic} can be deduced to be
\begin{equation}\label{equ: Weingarten asymp}
    \Wg(\sigma^{-1}\tau, m)= \begin{cases}
m^{-k}(1+\mathcal{O}(\frac{1}{m})) & \sigma=\tau \\
\mathcal{O}(m^{-(k+1)}) & \text{otherwise.}
\end{cases}
\end{equation}

\begin{theorem}\label{thm: MP}
Let $\rho\sim \Gre (n,d_A;d_B)$, and assume that as $d_A,d_B\rightarrow \infty$,  $\frac{d_A}{d_B} \rightarrow c \in (0,\infty)$. Then the empirical spectral measure
$\frac{1}{d_{A}}\sum_{i=1}^{d_{A}}\delta_{\lambda_{i}(d_{A}\rho)}$ converges in probability to the Marchenko-Pastur law $\mu^{\text{MP}}_{c/n}$. 
    
\end{theorem}
\begin{proof}
We begin by computing the asymptotics of the tracial moments. Consider the formula for moments from Theorem \ref{thm: tracial moments}:
\begin{equation*}
\ex\left[\frac{1}{d_{A}}\Tr(d_{A}\rho)^{k}\right]=\sum\limits_{\sigma, \tau \in \Sym_k}\Wg(\sigma^{-1}\tau,d_Ad_B)d_A^{\#(\gamma^{-1}\sigma)+k-1}d_B^{\#(\sigma)}n^{\#(\tau)-k}.
\end{equation*}
Recall the triangle inequality $\#(\pi^{-1}\gamma_n)+\#(\pi)\leq n+1$, with equality precisely for non-crossing permutations. Using this inequality and the Weingarten estimates, we may write the tracial moments as follows:
 \[
\begin{aligned}
\ex\left[\frac{1}{d_{A}}\Tr(d_{A}\rho)^{k}\right]
&= \sum_{\sigma\in \Sym_k}
   (d_A d_B)^{-k}
   d_A^{\#(\gamma\sigma^{-1})+k-1}
   d_B^{\#(\sigma)}
   n^{\#(\sigma)-k}\left(1+\mathcal{O}\left( d_{A}^{-1}\right)\right) + \mathcal{O}(d_{A}^{-2})
   \\
&= \sum\limits_{\sigma \in NC(k)} c^{1-\#(\sigma)}d_B^{\#(\sigma)-k}n^{\#(\sigma)-k}+ \mathcal{O}(d_{A}^{-2}).
\end{aligned}
\]

After scaling by $d_A$ and using $d_A/d_B=c$, we get:

\[
\begin{aligned}
\lim_{d_{A},d_{B}\rightarrow \infty}    \ex\left[\frac{1}{d_A}\Tr(d_A\rho)^k\right]&=\left(\frac{d_A}{d_B}\right)^{k-1}\sum\limits_{\sigma \in NC(k)}c^{1-\#(\sigma)}n^{\#(\sigma)-k} \\
    &=\sum\limits_{\sigma \in NC(k)}\left(\frac{c}{n}\right)^{k-\#(\sigma)}\\
    &= M_k(\mu_{MP}).
\end{aligned}
\]
This formula for the Marchenko-Pastur moments can be found as equation (4.18) in \cite{mingo2017free}. 

The sequence of measures $\mu_{d\rho}$ is tight since taking $\epsilon = \frac{1}{R}$, for $R>0$ , we have that: 
\begin{equation*}
    \sup_{d_{A}} \mu_{d_{A} \rho}( (R,\infty)) \leq \frac{1}{R}\int x d\mu_{d_{A} \rho}(x) = \frac{1}{R} = \epsilon.
\end{equation*}
Thus, we have that any sequence of our measures have weakly convergent subsequences. We may also bound the tail of all higher moments
\begin{equation*}
    \int_{x> R}x^{\ell} d\mu_{d_{A}\rho}(x) \leq \frac{1}{R}\int x^{\ell+1}d\mu_{d_{A}\rho}(x).
\end{equation*}
This implies that along a subsequence on which all moments converge, the right-hand side is bounded, so as a result moments pass to the weak limit. The Marchenko-Pasutr law is compactly supported so it is determined by its moments. Hence, we have shown convergence in distribution.

In order to prove convergence in probability, it suffices to bound the variance and use Chebyshev's inequality. To bound the variance, first note that 
\[
\ex\left[\Tr(\rho^k)^2\right] =\sum_{i_1,\dots ,i_k,i_{k+1},\dots ,i_{2k}}\ex\left[\rho_{i_1,i_2}\rho_{i_2,i_3}\dots\rho_{i_k,i_1}\rho_{i_{k+1},i_{k+2}}\dots\rho_{i_{2k},i_{k+1}}\right].
\]
Using Proposition \ref{prop: frommatsumoto2023moments}, we can specialize it to the case that we have the product of two traces of powers: \begin{align*}
\ex\left[\Tr(\rho^k)^2\right]&=\sum_{\sigma,\tau\in \Sym_{2k} }\sum_{\mathbf{i}}\delta_{\sigma}(\mathbf{i}, \gamma_{k,k}({\mathbf{i}}))\,d_B^{\#(\sigma)}\Wg(\sigma^{-1}\tau,m)\,n^{\#(\tau)-2k}\\
&=\sum_{\sigma,\tau \in \Sym_{2k} }d_A^{\#(\gamma_{k,k}\sigma^{-1})}d_B^{\#(\sigma)}\Wg(\sigma^{-1}\tau,m)\,n^{\#(\tau)-2k}.
\end{align*}
Applying the leading-order asymptotics once more, we find that
\[
\ex\left[\Tr(\rho^k)^2\right]=\sum_{\sigma\in \Sym_{2k}}d_A^{\#(\gamma_{k,k}\sigma^{-1})}d_B^{\#(\sigma)}\, m^{-2k}\, n^{\#(\sigma)-2k}+o(1).
\]
Likewise, after using leading-order asymptotics for the tracial moment squared, we have that
\begin{align*}
\ex\left[\Tr(\rho^k)\right]^2&=\sum_{\sigma, \tilde{\sigma} \in \Sym_k }m^{-2k}\,d_A^{\#(\gamma \sigma^{-1})+\#(\gamma \tilde{\sigma}^{-1})}\,d_B^{\#(\sigma)+\#(\tilde{\sigma})}n^{\#(\sigma)+\#(\tilde{\sigma})-2k}+o(1).
\end{align*}
Observe that, viewing $\tilde{\sigma}$ as a permutation on $\{k+1,\dots ,2k\}$, we can instead sum over permutations  $\sigma \times\tilde{\sigma}\in \Sym_k \times \Sym_k \subset \Sym_{2k}$, and we have 
\[
\#(\gamma \sigma^{-1})+\#(\gamma \tilde{\sigma}^{-1})=\#(\gamma_{k,k}(\sigma\times\tilde{\sigma})^{-1}).
\]
Hence, our variance to leading order is

\begin{align*}
\operatorname{Var}({\Tr(\rho^k)})&=\sum_{\pi \in \Sym_{2k} \setminus (\Sym_k\times \Sym_k)}d_A^{\#(\gamma_{k,k}\pi^{-1})}\, d_B^{\#(\pi)}\,m^{-2k}\,n^{\#(\pi)-2k} + o(1)\\
&= \sum_{\pi \in \Sym_{\text{NC}}(p,q)}d_A^{2k-\#(\pi)}\, d_B^{\#(\pi)}\,m^{-2k}\,n^{\#(\pi)-2k}+o(1).
\end{align*} 

Now, if we rescale by $d_A$ and use $\frac{d_A}{d_B}=c$, we obtain
\[
\operatorname{Var}\left(\frac{1}{d_A}\Tr((d_A\rho)^k\right)=\frac{M_k}{d_A^2}(1+o(1))\rightarrow 0
\]
as $d_A,d_B \rightarrow\infty$. Note that $M_k$ depends only on $c$ and $k$. Thus, applying Chebyshev's inequality, we find that for any $\epsilon>0$ and any positive integer $k$ we have that
\begin{equation*}
     \mathbb{P}\left(\left|\frac{1}{d_A}\Tr((d_A\rho)^k -\int x^{k} \mu^{\text{MP}}_{c/n}(x) \right| >\epsilon\right) \leq \frac{1}{\epsilon^2}\operatorname{Var}({\Tr(\rho^k)}).
\end{equation*}
Since all measures involved are compactly supported, the above steps imply that for any test function $\varphi \in C_{0}^{1}(\mathbb{R})$,
\begin{equation*}
     \mathbb{P}\left(\left|\int \varphi d\mu_{d_{A}\rho} -\int \varphi \mu^{\text{MP}}_{c/n} \right| >\epsilon\right) \rightarrow 0,
\end{equation*}
as $d_{A},d_{B} \rightarrow \infty$.
 


\end{proof}

An alternative asymptotic regime is obtained by fixing the dimension of one Hilbert space and letting the dimension of the other tend to infinity. In this case, Grassmannian density matrices converge to the normalized identity operator, which is the maximally mixed state.

\begin{prop}\label{prop: identity}
Fix $d_A$, and set $\frac{n}{d_{B}}=c\in(0,\infty)$. Then any sequence of states $\rho\sim\Gre(n,d_A;d_B)$ converges in probability to $\frac {1} {d_A} I_{d_A}$ as $d_{B} \rightarrow \infty$.
\begin{proof}
We use the leading-order asymptotic estimate for the Weingarten function as before to find that, for large $d_B$,
\begin{align*}
\ex\left[\prod_{t=1}^k\rho_{i_t,j_t}\right]&= \sum_{\sigma \in \Sym_k}\delta_{\sigma}(\mathbf{i},\mathbf{j})\, d_B^{\#(\sigma)}\,(d_Ad_B)^{-k}\, (c d_B)^{\#(\sigma)-k}(1+o(1))\\
&=\sum_{\sigma \in \Sym_k}\delta_{\sigma}(\mathbf{i},\mathbf{j})\,d_A^{-k}\, d_B^ {2(\#(\sigma)-k)}c^{\#(\sigma)-k}(1+o(1))\\
&= \delta_{id.}(\mathbf{i},\mathbf{j})\,d_A^{-k}(1+o(1)).
\end{align*}
So in the limit we have that
\[
\ex\left[\rho_{i,j}\right]\rightarrow \frac{1}{d_A}\delta(i,j).
\]
Additionally, we find that
\begin{align*}
\operatorname{Var}(\rho_{i,j})&=\ex\left[\rho_{i,j}^2\right]-\ex\left[\rho_{i,j}\right]^2\\
&= \delta(i,j)\,d_A^{-2}(1+o(1))-\delta(i,j)\,d_A^{-2}(1+o(1))\\
&= o(1).
\end{align*}
Another application of Chebyshev's inequality gives the result.
\end{proof}
\end{prop}

Theorem \ref{thm: MP} and Proposition \ref{prop: identity} generalize Theorem 5 and Corollary 2 of \cite{nechita2007asymptotics}, respectively, which were for the $n=1$ case.

\subsection{Asymptotic freeness}    

In this subsection, we prove some properties of the reduced density operators of the Grassmannian ensemble in the language of free probability theory. Additionally, we obtain an alternative proof of Theorem \ref{thm: MP}. First, we will give some brief background on free probability theory. 

    \begin{definition}
        A \textit{non-commutative probability space} consists of a unital algebra $\mathcal{A}$ and a linear map $\varphi: \mathcal{A\rightarrow}\mathbb{C}$ such that $\varphi(1_{\mathcal{A}}) = 1$. An element  $a\in \mathcal{A}$ is called a \textit{non-commutative random variable}.  
    \end{definition} 
    
     Given a non-commutative random variable $a$, one is interested in its moments $\{\varphi(a^{\ell})\}_{\ell\geq 1}$. In this context, the moments characterize the distribution of the non-commutative random variable.
     \begin{definition}
         Consider a sequence of non-commutative probability spaces $(\mathcal{A}_{N},\varphi_{N})_{N}$ and another non-commutative probability space $(\mathcal{A},\varphi)$. Let $(a_{N})_{N}$ be a sequence of non-commutative random variables where $a_{N}\in \mathcal{A}_{N}$, and let $a\in \mathcal{A}$. Then we say $a_{N}$ converges in distribution to $a$, denoted $a_{N}\xrightarrow{\text{dist.}} a$ if 
         \begin{equation*}
             \lim_{N \rightarrow\infty}\varphi(a_{N}^{\ell}) = \varphi(a^{\ell})
         \end{equation*}
         for any $\ell \in \mathbb{N}$.
     \end{definition}

More generally, given a set of non-commutative random variables, we are interested in their mixed moments. In the commutative case, the independence of random variables implies the factorization of the expectation values of mixed moments. With non-commutative random variables, the analogous notion of independence is freeness and was first considered by Voiculescu \cite{voiculescu2006symmetries}. In our work, we are specifically interested in the notion of asymptotic freeness of random matrices. 
     
     A powerful definition of freeness, first given by Speicher \cite{speicher1994multiplicative}, is that all mixed free cumulants of elements of the subalgebra are zero. In free probability theory, freeness is analogous  to the notion of independence in classical probability theory, in which the vanishing of mixed (classical) cumulants is equivalent to independence. Free cumulants of a non-commutative random variable $a$ can be defined recursively via the relations 
\begin{equation*}
    \varphi(a^{\ell}) = \sum_{\pi \in NC(\ell)} \prod_{B \in \pi}\kappa_{|B|}(a),
\end{equation*}
where the product is over blocks $B_{i}$ of the partition $\pi$. The inverse formula can be found via M\"obius inversion: 
\begin{equation*}
    \kappa_{\ell}(a) = \sum_{ \pi \in NC(\ell)} \mu_{NC(\ell)}(\pi, 1_{\ell}) \prod_{B \in \pi}\varphi(a^{|B|}),
\end{equation*}
where the M\"obius function $\mu_{NC(\ell)}(\pi, 1_{\ell})$ is a product of signed Catalan numbers.  
\begin{definition}
    The subalgebras $\mathcal{A}_1,...,\mathcal{A}_{r}$ are free if and only if all mixed cumulants vanish.
\end{definition}

     We are particularly interested in the example in which the unital algebra is the algebra of random $N$-by-$N$ matrices and the linear map is the expectation value of the normalized trace. Just as in the case of commutative random variables, the set of all such random matrices forms an algebra. A relevant example of a random non-commutative variable is then the induced ensemble. As an example, Theorem \ref{thm: MP} implies that given a sequence of induced ensembles of $\rho\sim \mathcal{GE}(n,d_{A};d_{B})$, such that $\frac{d_{A}}{d_{B}}=c$, 
     \begin{equation}
\rho \xrightarrow{\text{dist.}} s^{\textit{MP}},
     \end{equation}
     as $d_{A},d_{B}\rightarrow \infty$ where $s^{\textit{MP}}$ is the Marchenko-Pastur random variable with parameter $c$.

Consider the set of orthogonally induced states $(\rho_{t})_{t}$ on $\mathcal{H}_{A}$ as a subsystem of $\mathcal{H}_{A}\otimes \mathcal{H}_{B}$. Theorem \ref{thm: Mixed tracial moments} gives us a method to compute mixed moments of these random states. The following asymptotic formula for moments follows from Theorem \ref{thm: Mixed tracial moments} and will be used to prove the asymptotic freeness of these states.
\begin{lemma}\label{cor: limiting moments}
 Let $\frac{1}{n}\sum_{t=1}^{n} \rho_{t} \sim \Gre(n,d_{A};d_{B})$. Provided that $\frac{d_{A}}{d_B} \rightarrow c\in  (0,\infty)$, we have that
    \begin{equation*}
        \lim_{d_{A},d_{B} \rightarrow \infty} d_{A}^{\ell-1}\ex[\Tr\rho_{i_{1}}^{\ell_{1}},\rho_{i_{2}}^{\ell_{2}},\dots ,\rho_{i_{k}}^{\ell_{k}} ] = \sum_{\tau \in  \text{NC}(\ell)\cap \stab(w)}c^{\ell-\#(\tau)},
    \end{equation*}
    where neighbouring states $\rho_{i_{j}}$ are distinct. The sum is over all non-crossing permutations that respect the colours of the word $w$; that is, $i_{r} = i_{p}$ when $\pi(i_{r}) =i_{p}$.
\end{lemma}

\begin{proof}
    Using Theorem \ref{thm: Mixed tracial moments} for mixed moments and the leading-order asymptotics for the Weingarten function 
\[
\Wg(\sigma, m)= \begin{cases}
m^{-\sum_{i=1}^{k}\ell_{i}}(1+\mathcal{O}(m^{-1})) & \sigma=\text{id} \\
\mathcal{O}(m^{-(\sum_{i=1}^{k}\ell_{i}+1)}) & \text{otherwise},
\end{cases}
\]
we can write 
    \begin{align*}
        \ex[\Tr\rho_{i_{1}}^{\ell_{1}}\rho_{i_{2}}^{\ell_{2}}\dots \rho_{i_{k}}^{\ell_{k}}] &= \sum_{\substack{\sigma\in \Sym_{\ell_{1}+...+\ell_{k}}\\ \tau \in \stab(w)}}W_{g}(\sigma,m) d_{A}^{\#(\gamma_{\ell}\sigma\tau)}d_{B}^{\#(\sigma \tau)}\\
        &= \sum_{ \tau \in \stab(w)} m^{-\ell}(1+\mathcal{O}(m^{-1}))d_{A}^{\#(\gamma_{\ell} \tau)}d_{B}^{\#(\tau)} 
        + \mathcal{O}(m^{-(\ell+1)}).
    \end{align*}
   The permutations that maximize the power of $d_{A}$ are non-crossing permutations that are in the stabilizer group. Thus, we can substitute $\#(\gamma\tau) = \ell+1-\#(\tau)$ into our sum. Multiplying by $d_{A}^{\ell-1}$, one for each instance of $\rho_{i}$ and one to normalize the trace, then taking the limit gives us our result.
\end{proof}

\begin{theorem}\label{thm: freeness}
  Let $\frac{1}{n}\sum_{t=1}^{n} \rho_{t} \sim  \Gre(n,d_{A};d_{B})$. Then $d_{A}\rho_{t}\xrightarrow{\text{dist.}} s_{t}$ as $d_{A}$ and $d_{B}$ go to infinity such that $\frac{d_{A}}{d_{b}}=c \in (0,\infty)$,
    where $\{s_{t}\}_{t=1}^{n}$ are free Marchenko-Pastur random variables with parameters $c$ and $n$.
\end{theorem}
\begin{proof}
   If $\{s_{t}\}_{t=1}^{n}$ are free Marchenko-Pastur random variables with parameters $c$ and $n$, then the free cumulant of any word $w$ over the alphabet $\{\rho_{1},...,\rho_{n}\}$ is zero except when the word is only composed of one letter, i.e., $w=\rho_{i}^{\ell}$, in which case $\kappa_{\ell}(\rho_{i}^{\ell}) = (c/n)^{\ell-1}$. Substituting this formula into the moment-free cumulant formula produces a sum over non-crossing permutations such that each block contains only one state label. This is precisely the formula from Lemma \ref{cor: limiting moments}. 
\end{proof}

A particularly useful consequence of Speicher's result is that, for two random matrices $A$ and $B$ that are asymptotically free, the free cumulants of $A+B$ are the sum of the free cumulants of $A$ and the free cumulants of $B$. In terms of generating functions:
\begin{equation*}
    R_{A+B}(z) = \sum_{n=0}^{\infty} \kappa^{A+B}_{n+1}z^{n} = \sum_{n=0}^{\infty} \kappa^{A}_{n+1}z^{n} + \sum_{n=0}^{\infty} \kappa^{B}_{n+1}z^{n} = R_{A}(z) + R_{B}(z).
\end{equation*}
Using this result, we now present an alternative proof of Theorem \ref{thm: MP}, but for convergence in distribution.
\begin{proof}
 Let $(\rho_{t})_{t}\sim  \Gre(n,d_{A};d_{B})$. As discussed above, we know $\kappa(\rho_{i}^{\ell}) = c^{\ell -1}$, so 
\begin{equation*}
    R_{d_{A}\rho_{i}}(z) = \sum_{\ell=1}^{\infty}c^{\ell -1} z^{\ell-1} = \frac{1}{1-cz}. \end{equation*}
 By Theorem \ref{thm: freeness} we have that 
    \begin{equation*}
        R_{d_{A}\rho}(z) = \frac{1}{n}\sum_{i=1}^{n}R_{d_{A} \rho_{i}}(z/n) = \sum_{\ell=1}^{\infty} \left( \frac{c}{n}\right)^{\ell-1}z^{\ell-1}. \end{equation*}
     Since our distribution is compactly supported, it is uniquely determined by its moments and therefore free cumulants. 
\end{proof}

\subsection{The average entropy}

The entropy of a quantum system was first introduced by von Neumann as a measure of statistical uncertainty in the system. 
The \textit{von Neumann entropy} generalizes the  entropy from other  
fields, in particular the Shannon entropy from classical information theory, to quantum systems. 
The von Neumann entropy of a quantum system described by a density operator $\rho$ is defined as  
\begin{equation*}
    S(\rho)=-\Tr(\rho \log \rho).
\end{equation*}
Note that the function $f(x)=-x\log x$ is an analytic function on non-negative real $x$, and on this domain, it is concave down everywhere. 
The functional calculus can be used to define $f(\rho)$.
The trace of the operator $f(\rho)$ will depend only on the spectrum of $\rho$, which consists of the eigenvalues $\{\lambda_j\}$ of $\rho$. 
Indeed, it can be written as 
$$S(\rho)=-\sum_{j} \lambda_j\log \lambda_j.$$
The entropy $S(\rho)$ is nonnegative and bounded above by the logarithm of the dimension of the system.

In quantum information, entropy can also be used as a measure of entanglement of states on bipartite systems.
The  state of a bipartite system $\mathcal{H}_A\otimes \mathcal{H}_B$ described by the density operator $\rho_{AB}$ is ``more'' entangled if the entropy of the reduced density operator on a subsystem is higher. 
The quantity $S(\Tr_B(\rho_{AB}))$ is called \textit{entanglement entropy} and can be averaged over an ensemble of random states. As mentioned in the introduction, Page conjectured the now-celebrated formula for the average entropy of the induced ensemble. We will use Theorem \ref{thm: MP} to compute the entropy in the scaled limit.
\begin{corollary}\label{cor: entropy}
Let $\rho\sim \Gre (n,d_A;d_B)$. Under the assumptions of Theorem \ref{thm: MP}, we have that
$$ \ex[(\rho)]=\log d_{A} -\begin{cases}\log\left(\frac{c}{n}\right) +\frac{n}{2 c} & c\geq n \\ &\\ \frac{c}{2n} & c\leq n \end{cases} +o(1).$$
\end{corollary}
\begin{proof}
Note that the entropy function is unbounded so weak convergence is insufficient to pass to the limit. We require a tail estimate, so consider $0 \leq x \log x \leq \frac{\log R}{R}x^2 $ for any $x\geq R \geq e$, where
\begin{equation*}
  \mathbb{E} \left[ \int_{x >R}x\log x d\mu_{d\rho}(x) \right]\leq \frac{\log R}{R} \mathbb{E}\left[ \int x^2 d\mu_{d_{A} \rho}(x)\right].
\end{equation*}
The second moment \eqref{equ: secondtracialmoment} is finite in $d_{A}$ and $d_{B}$, so as $R$ goes to infinity we have that 
$$\mathbb{E} \left[ \int x\log x d\mu_{d\rho}(x) \right]\rightarrow\int x\log x\, d\mu_{c/n}^{\text{MP}}(x).$$

    We wish to evaluate 
    \begin{align*}
        \mathbb{E}[\rho \log \rho] &= \log d_{A} -\int x\log x\, d\mu_{c/n}^{\text{MP}}(x)\\
        &= \log d_{A}-\frac{n}{c}\int_{a}^{b}\frac{\sqrt{(b-x)(x-a)}}{2 \pi  x} dx,
    \end{align*}
 since the delta function does not contribute anything. Denote the remaining integral   as $I$. We apply the substitution $x = 1+ c/n + 2 \sqrt{c/n}\cos(\theta)$ for $\theta \in [0,\pi]$ to arrive at 
    \begin{equation*}
        I = \frac{2}{\pi} \int_{0}^{\pi} \log (1 + c/n + 2 \sqrt{c/n} \cos( \theta)) \sin(\theta)^2 d\theta.
    \end{equation*}
    
    We have two possibilities to consider. First, if $c/n \in (0,1)$, then $1 + c/n + 2 \sqrt{c/n} \cos( \theta) = (1+ \sqrt{c/n} e^{i \theta})(1+ \sqrt{c/n} e^{-i \theta}) $, allowing us to expand the logarithm uniformly as 
    \begin{equation*}
        \log (1 + c/n + 2 \sqrt{c/n} \cos( \theta)) = 2 \sum_{j=1}^\infty \frac{(-1)^{j+1}(c/n)^{j/2}}{j} \cos(j \theta).
    \end{equation*}
    The relevant integral identity is now 
    \begin{align*}
        \int_{0}^{\pi} \cos( j \theta) \sin (\theta)^2 d\theta &= \frac{1}{4}
         \int_{0}^{\pi} \cos( j \theta) (1 - \cos(2 \theta))^2 d\theta
    \end{align*}
    which is equal to $-\pi/4$ when $j=2$ and zero otherwise. By the orthogonality of cosine, only the $\cos( 2 \theta)$ coefficients contribute when the integrand is expanded.  The result is $I =c/2n$.
    
Alternatively, if $c/n >1$, then we need to factor out our constant to apply the expansion i.e. write
\begin{equation*}
    \log\left( 1 + c/n + 2\sqrt{c/n} \cos (\theta) \right) = \log (c/n) + \log\left( 1 + n/c + (2n/c) \cos( \theta)\right).
\end{equation*}
Applying the same expansion procedure as above, we have that the integral coming from the left hand term evaluates to $\log (c/n)$ and the right-hand term evaluates to $n/2c$.
\end{proof}

\section*{Appendix: Numerical simulation}

\begin{figure*}[t]
    \centering
    \begin{subfigure}[t]{0.49\linewidth}
        \centering
        \includegraphics[width=\linewidth]{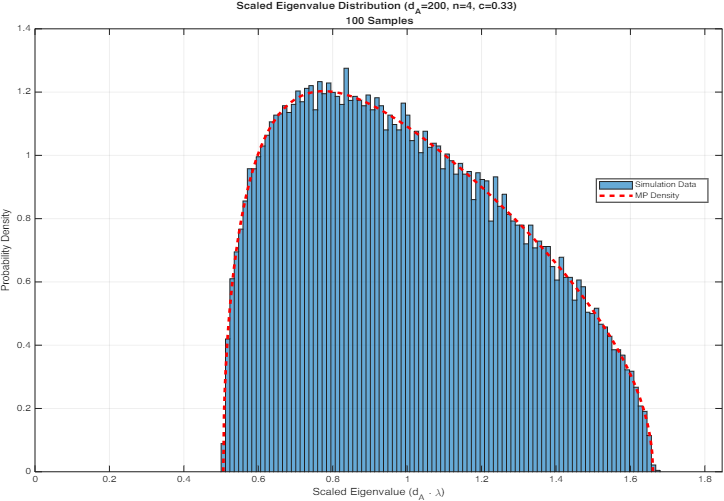}
    \end{subfigure}%
    \hfill
    \begin{subfigure}[t]{0.49\linewidth}
        \centering
        \includegraphics[width=\linewidth]{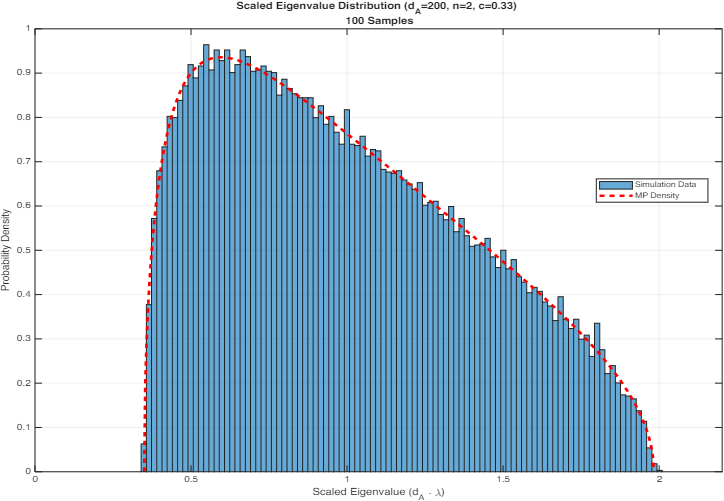}
    \end{subfigure}
    \caption{The empirical eigenvalue distributions of $d_A\rho$ for $d_A=200$ and $c=0.33$, with $n=4$ on the left and $n=2$ on the right, obtained using Algorithm~\ref{alg:rmt_simulation}. The red curve in each panel is the Marchenko-Pastur density with parameter $c/n$.}
    \label{fig:eigenvalue_distributions}
\end{figure*}

To obtain the eigenvalue distributions plotted in 
Figure~\ref{fig:eigenvalue_distributions}, we performed a Monte Carlo 
simulation over $N$ random subspaces, as summarised in 
Algorithm~\ref{alg:rmt_simulation}.

\begin{algorithm}
\caption{Monte Carlo estimation of the reduced density matrix eigenvalue distribution}
\label{alg:rmt_simulation}
\begin{algorithmic}[1]

\REQUIRE Subsystem dimensions $d_A,\, d_B$; subspace dimension $n$; 
         number of trials $N$
\ENSURE  Empirical distribution of scaled eigenvalues $d_A \lambda$

\FOR{$i = 1$ \TO $N$}
    \STATE Sample a random $n$-dimensional subspace of 
           $\mathbb{C}^{d_A \otimes d_B}$ via QR decomposition of a 
           Gaussian random matrix
    \STATE Reshape each column $V_{:,k}$ of $Q$ into a matrix 
       $M_k \in \mathbb{C}^{d_A \times d_B}$, \; $k = 1, \dots, n$
    \STATE Compute the reduced density matrix of subsystem $A$:
           \[
               \rho_A \;=\; \frac{1}{n}\sum_{k=1}^{n} M_k M_k^\dagger,
               \qquad M_k \in \mathbb{C}^{d_A \times d_B}
           \]
    \STATE Record the scaled eigenvalues $\{d_A\,\lambda_j\}_{j=1}^{d_A}$ 
           of $\rho_A$
\ENDFOR

\STATE Plot the histogram of all scaled eigenvalues and 
       overlay the Marchenko--Pastur density with aspect ratio 
       $c = d_A/(n d_B)$

\end{algorithmic}
\end{algorithm}

Using this algorithm, we computed the entropy for various small values of the parameters $n,d_{A},$ and $d_{B}$ in Table \ref{tab:my-table}. When $n=1$, we can compare the simulated value with the known value in Page's formula.

\begin{table}[H]
\centering
\begin{tabular}{lllll}
\hline
 $n$ & $d_{A}$&$d_{B}$ &  Numerical estimate & Known value \\ \hline
 \multicolumn{1}{|l|}{1}& \multicolumn{1}{l|}{2} & \multicolumn{1}{l|}{2} &  \multicolumn{1}{|l|}{0.333140}  & \multicolumn{1}{l|}{0.333333} \\ \hline
 \multicolumn{1}{|l|}{1} & \multicolumn{1}{l|}{2} & \multicolumn{1}{l|}{3} &   \multicolumn{1}{|l|}{0.449938} & \multicolumn{1}{l|}{0.45} \\ \hline
  \multicolumn{1}{|l|}{1} & \multicolumn{1}{l|}{3} & \multicolumn{1}{l|}{3} &  \multicolumn{1}{|l|}{0.66221} & \multicolumn{1}{l|}{0.66230} \\ \hline
  \multicolumn{1}{|l|}{1} &  \multicolumn{1}{l|}{3} & \multicolumn{1}{l|}{4} & \multicolumn{1}{|l|}{0.769829} & \multicolumn{1}{l|}{0.769877} \\ \hline
\multicolumn{1}{|l|}{1}& \multicolumn{1}{l|}{4} & \multicolumn{1}{l|}{4} & \multicolumn{1}{|l|}{0.922409} & \multicolumn{1}{l|}{0.922396}\\ \hline
 \multicolumn{1}{|l|}{2} & \multicolumn{1}{l|}{2} & \multicolumn{1}{l|}{2} & \multicolumn{1}{|l|}{0.5853667}  \\ \cline{1-4}
 \multicolumn{1}{|l|}{2} & \multicolumn{1}{l|}{3} & \multicolumn{1}{l|}{3} & \multicolumn{1}{|l|}{0.912718}  \\ \cline{1-4}
 \multicolumn{1}{|l|}{2} & \multicolumn{1}{l|}{3} & \multicolumn{1}{l|}{4} & \multicolumn{1}{|l|}{0.952890}  \\ \cline{1-4}
 \multicolumn{1}{|l|}{3} & \multicolumn{1}{l|}{3} & \multicolumn{1}{l|}{3} & \multicolumn{1}{|l|}{0.994017}  \\ \cline{1-4}
 \multicolumn{1}{|l|}{3} & \multicolumn{1}{l|}{3} & \multicolumn{1}{l|}{4} & \multicolumn{1}{|l|}{1.012131}  \\ \cline{1-4}
\end{tabular}
\caption{Numerical values of the entropy of the Grassmannian ensemble for various values of the parameters $n,d_{A},$ and $d_{B}$ obtained using Algorithm \ref{alg:rmt_simulation}. The rightmost column contains the known values from Page's formula when $n=1$.}
\label{tab:my-table}
\end{table}

\section*{Acknowledgements}
We would like to thank the Fields Institute for organizing the Fields Undergraduate Summer Research Program in 2025. Liam Hickey, Allison Klingler, Andrew Parks, and Aryan Gadhave were participants in this program while working on this paper. The project was proposed and
supervised by Asghar Ghorbanpour, Masoud Khalkhali, and Nathan Pagliaroli. Nathan Pagliaroli would like to thank James Mingo for helpful discussions and for hosting him at Queen's University during the completion of this paper. The authors also thank the Natural Sciences and Engineering Research Council of Canada (NSERC) for financial support. ChatGPT 6.1 Sol was used to proofread this article and provide corrections to arguments such as the computation of the entropy in Corollary \ref{cor: entropy}.



\printbibliography

\end{document}